\documentstyle[a4,epsf,amsfonts,amssymb]{article} 

\newcommand{\vs}{\vspace*{5mm}}  

\def\ben{\begin{equation}}
\def\een{\end{equation}}
\def\bena{\begin{eqnarray}}
\def\eena{\end{eqnarray}}

\newcommand{\gD}{{}^{(\gamma)} \!{\nabla}}
\newcommand{\kD}{{}^{(k)} \!{\nabla}}

\newtheorem{theorem}     {Theorem}

\newtheorem{proposition} {Proposition} 
\newtheorem{definition}  {Definition}

\newtheorem{corollary}   {Corollary}
\newtheorem{remark}       {Remark}

\newcommand{\bproof}{\setlength{\parindent}{0mm}{\it Proof.{~~}}}
\newcommand{\eproof}{\hfill $\Box$\setlength{\parindent}{5mm}}

\begin{document}

\def\abstract#1
{\begin{center}{\large Abstract}\end{center}\par #1}
\def\title#1{\begin{center}{{\Large #1}}\end{center}}
\def\author#1{\begin{center}{{\large  #1}}\end{center}}
\def\address#1{\begin{center}{\it #1}\end{center}} 


\begin{flushright}
YITP-00-54 \\ 
DAMTP-2000-116 
\end{flushright}
\vskip 0.5cm

\title{ {\bf Convex Functions and Spacetime Geometry} }

\vskip 0.3cm

\author{ 
Gary W. Gibbons$^{\dag}$
\footnote{E-mail address: gwg1@amtp.cam.ac.uk }
and
Akihiro Ishibashi$^{\dag \ddag}$
\footnote{E-mail address: akihiro@yukawa.kyoto-u.ac.jp}
}

\vskip 0.3cm

\address{
$^\dag$ 
DAMTP, Center for Mathematical Sciences,\\ 
University of Cambridge, \\
Wilberforce Road, Cambridge CB3 0WA, the United Kingdom \\
}

\address{
$^\ddag$
Yukawa Institute for Theoretical Physics,\\
Kyoto University,\\
Kyoto 606-8502, Japan\\
}
                            

\vskip 0.3cm
\begin{abstract}
Convexity and convex functions play an important role
in theoretical physics. 
To initiate a study of the possible uses of convex functions 
in General Relativity, 
we discuss the consequences of a spacetime $(M,g_{\mu \nu})$ or 
an initial data set $(\Sigma, h_{ij}, K_{ij})$ admitting 
a suitably defined convex function. 
We show how the existence of a convex function on a spacetime 
places restrictions on the properties of the spacetime geometry. 
\end{abstract}

\section{Introduction}
\label{sect:introduction}

Convexity and convex functions play an important role
in theoretical physics. For example, Gibbs's approach to 
thermodynamics~[G] 
is based on the idea that the free energy 
should be a convex function. 
A closely related concept is that of a convex cone
which also has numerous applications to physics. 
Perhaps the most familiar example is the lightcone of Minkowski spacetime.
Equally important is the convex cone of mixed states
of density matrices in quantum mechanics. 
Convexity and convex functions also have important applications to geometry, 
including Riemannian geometry \cite{Udriste.C1994}. 
It is surprising therefore that, to our knowledge, that techniques making 
use of convexity and convex functions have played no great role 
in General Relativity. 
The purpose of this paper is to initiate a study of the possible
uses of such techniques. As we shall see, the existence of a convex
function, suitably defined, on a spacetime places important restrictions
on the properties of the spacetime. 
For example it can contain no closed spacelike geodesics.  
We shall treat Riemannian
as well as Lorentzian manifolds with an eye to applications 
in black hole physics, cosmology and quantum gravity. 
The detailed results and examples will be four-dimensional
but extensions to other dimensions are immediate.

One way of viewing the ideas of this paper is in terms of a sort of duality
between paths and particles on the one hand and functions and waves 
on the other. Mathematically the duality corresponds
to interchanging range and domain. A curve $x(\lambda)$ is a map 
$x: {\Bbb R} \rightarrow M$ while a function $f(x)$ 
is a map $f:M \rightarrow {\Bbb R}$. 
A path arises by considering invariance under
diffeomorphisms of the domain (i.e. of the world volume)
and special paths, for example geodesics have action functionals
which are reparametrization invariant. Much effort, physical and
mathematical has been expended on exploring the global
properties of spacetimes using geodesics. 
Indeed there is a natural
notion of convexity based on geodesics. 
Often a congruence of geodesics is used.

On the dual side, one might consider properties
which are invariant  under diffeomorphisms $f(x)\rightarrow g(f(x))$
of the range or target space. That is one 
may explore  the global properties
of spacetime using the foliations provided
by the level sets of a suitable function. The analogue
of the action functional for geodesics is one like
\ben
\int_M \sqrt {\epsilon \nabla _\mu f \nabla^\mu f}, 
\een
where $\epsilon =\pm$, depending upon whether $\nabla _\mu  f$ is
spacelike or timelike, and which is invariant under 
reparametrizations of the range. 
The level sets  of the solutions of the Euler-Lagrange equations
are then minimal surfaces. One may also consider
foliations by totally umbilic surfaces or by ``trace $K$ equal constant" 
foliations, and this is often done in numerical relativity.
The case of a convex function then corresponds 
to a foliation by totally expanding hypersurfaces, that is, 
hypersurfaces with positive definite second fundamental form.

Another way of thinking about spacetime is in
terms of the causal structure provided by the Lorentzian
metric. However what we shall call a spacetime convex
function has a Hessian with Lorentzian signature
which also defines a causal structure, the light cone of which
always lies inside the usual spacetime lightcone. 
The spacetimes admitting spacetime convex functions 
have particular types of causal structures.

The organisation of the paper is as follows. 
In the next section, we consider convex functions on Riemannian manifolds, 
which are regarded as spacelike submanifolds embedded in a spacetime 
or as Euclidean solutions of Einstein's equation.  
We see that the existence of convex functions of strictly or uniformly 
convex type is incompatible with the existence of closed minimal 
submanifolds. 
We also discuss the relation between the convex functions 
and Killing vector fields. 
In section~\ref{sect:spacetime}, we give two definitions for convex functions
on Lorentzian manifolds: a {\sl classical convex function} 
and a {\sl spacetime convex function}. 
In particular, the latter defines a causal structure on a spacetime 
and has many important applications. 
We discuss the connection between 
spacetime convex functions and homothetic Killing vector fields. 
We show that, if a spacetime admits a spacetime convex function, then 
such a spacetime cannot have a marginally inner and outer trapped surface. 
In section~\ref{sec:cosmology}, we examine in what case a cosmological 
spacetime admits a spacetime convex function. 
We also examine the existence of a spacetime convex function on 
specific spacetimes, de-Sitter, anti-de-Sitter, and black hole spacetimes, 
in the subsequent sections~\ref{sec:deS},~\ref{sec:AdS}, and~\ref{sec:BH}, 
respectively. 
In section~\ref{sec:Levelsets}, 
we discuss the level sets of convex functions and foliations. 
Then, in section~\ref{sec:barries}, we consider 
constant mean curvature foliations, barriers, and convex functions. 
Section~\ref{sec:summary} is devoted to a summary.

\section{Convex functions on Riemannian manifolds} 
\label{sect:Riemannian}

\subsection{Definitions and a standard example}

We shall briefly recapitulate standard definitions 
of convex functions on an $n$-dimensional Riemannian manifold 
$(\Sigma,h_{ij})$. 
What follows immediately is applicable to any Riemannian manifold  
of dimension $n$ including one obtained as a spacelike submanifold 
of an $(n+1)$- dimensional spacetime.

\begin{definition}{\bf Convex function:} \\
A $C^\infty$ function $\tilde{f}:\Sigma \rightarrow {\Bbb R}$ is 
said to be {\em (strictly) convex} if the Hessian is positive semi-definite 
or positive definite, respectively 
\bena  
&& (i) \quad  D_i D_j \tilde{f} \ge 0 \; ,   
\\&&
(ii) \quad  D_i D_j \tilde{f} > 0 \;, 
\eena 
where $D_i$ is the Levi-Civita connection of the positive definite metric 
$h_{ij}$. 
\end{definition}

Properties of convex functions on Riemannian manifolds have been extensively 
studied~\cite{Udriste.C1994}.

\vs 
\noindent 
{\bf Convex Cones} \\
\noindent 
The set of convex functions on a Riemannian manifold $(\Sigma,h)$ 
forms a {\em convex cone} $C_\Sigma$, 
that is, an open subset of a vector space which is a cone: 
$\tilde{f} \in C_\Sigma \Rightarrow \lambda \tilde{f} \in C_\Sigma$, 
$\forall \,\lambda \in {\Bbb R}_+$ and which is convex: 
$\tilde{f}, \tilde{f}' \in C_\Sigma \Rightarrow \lambda 
\tilde{f} + (1-\lambda) \tilde{f}' \in C_\Sigma$, 
$\forall \, \lambda \in (0,1)$. 

\vs 
Suppose $\Sigma$ is a product $\Sigma = \Sigma_1 \times \Sigma_2$ 
with metric direct sum $ h = h_1 \oplus h_2$. 
Then it follows that $\tilde{f} = \tilde{f}_1 + \tilde{f}_2 \in C_\Sigma$, 
for any $\tilde{f}_1 \in C_{\Sigma_1},\; \tilde{f}_2 \in C_{\Sigma_2}$. 
This defines a direct sum of convex cones, 
\ben 
   C_\Sigma = C_{\Sigma_1} \oplus C_{\Sigma_2} \;. 
\een 

\begin{definition}{\bf Uniformly convex function:} \\
A smooth function $\tilde{f} : \Sigma \rightarrow {\Bbb R}$ 
is said to be {\em uniformly convex} 
if there is a positive constant $\tilde{c}$ such that 
\ben
(iii) \quad D_i D_j \tilde{f} \ge \tilde{c} h_{ij}, 
\een
i.e., 
$ \forall V \in T\Sigma , \; 
V^iV^j D_i D_j \tilde{f} \ge \tilde{c} h_{ij} V^iV^j $. 
\end{definition}

The standard example is an ${\Bbb E}^n$ with coordinates $\{ x^i \}$ and 
\ben
   \tilde{f} = \frac{1}{2} \left\{ (x^1)^2 + (x^2)^2 + \cdots 
                                   + (x^n)^2 \right\}. 
\een 

Note that the vector field $\tilde{D}^i := h^{ij}D_j \tilde{f}$ satisfies 
\ben 
   {\cal L}_{\tilde{D}} h_{ij} \ge 2 \tilde{c} \; h_{ij} \;. 
\een
If inequality is replaced by equality we have a homothetic 
conformal Killing vector field. 
In the example above, 
\ben
   \tilde{D} = x^1 \frac{\partial}{\partial x^1}
             + x^2 \frac{\partial}{\partial x^2} + \cdots 
             + x^n \frac{\partial}{\partial x^n} \;, 
\een
and $\tilde{D}$ is the standard dilatation vector field. 
Thus, $\Sigma$ admits a homothety. 

\vs 

\subsection{Convex functions and submanifolds} 
\label{subsec:submfds}

Suppose 
$(\Sigma,h_{ij})$ admits a convex function of the type  (ii), or (iii). 
Then we deduce the following propositions. 
\begin{proposition}
$\Sigma$ cannot be closed. 
\label{prop:not-closed}
\end{proposition}

\bproof
Otherwise by contraction (ii) or (iii) with $h^{ij}$ 
it would admit a non-constant subharmonic function.  
\eproof

\vs

\begin{proposition}
$(\Sigma,h_{ij})$ admits no closed geodesic curves. 
\label{prop:no-closed curve}
\end{proposition}

\bproof
Along any such a geodesic curve, we have 
\ben
 \frac{d^2 \tilde{f}}{ds^2} \ge \tilde{c} > 0 \;, 
\een
with $s$ being an affine parameter of the geodesic curve. 
This contradicts to the closedness of the geodesic curve. 
\eproof
\vs

\begin{proposition}
$(\Sigma,h_{ij})$ admits no closed minimal submanifold $S \subset \Sigma$. 
\label{prop:no-closed submfd}
\end{proposition}

\bproof
Let $k_{pq}$ ($p,q = 1,2, \cdot \cdot \cdot, n-1$), 
    $H_{pq}$, 
    $\kD_p$, and 
    $D_i$ 
be the induced metric on $S$, 
   the second fundamental form of $S$, 
   the Levi-Civita connection of $k_{pq}$, and 
   the projected Levi-Civita connection of $h_{ij}$, 
respectively. 
Then we have 
\ben
D_p D_q \tilde{f} = \kD_p \kD_q \tilde{f}
                        - H_{pq} \frac{\partial \tilde{f}}{\partial n} \;, 
\label{eq:Gauss}
\een
where $\partial \tilde{f} /\partial n = n^i \partial_i \tilde{f}$ with 
$n^i$ being the unit normal to $S$. 
Contracting with $h^{ij}$ and using the fact that $S$ is minimal, i.e., 
$k^{pq}H_{pq} = 0$, we find that $S$ admits a subharmonic function,
i.e.,
\ben
   \kD{}^2 \tilde{f} > 0 \;. 
\een
However if $S$ is closed this gives a contradiction. 
\eproof  
\vs 

As a simple example, which also illustrates
some of the subtleties involved,
consider the following metric, 
\ben
   h_{ij}dx^i dx^j = \left( 1 + \frac{M}{2 \rho} \right)^4 
                     (d\rho^2 + \rho^2 d\Omega^2) \;, \quad
M>0, \; \rho>0 \;. 
\een
This is the Schwarzschild initial data in isotropic coordinates. 
The two-surface $S$ defined as $\rho = M/2\; (r = 2M)$ is totally geodesic 
($H_{pq} = 0$) and hence minimal $(k^{pq}H_{pq} = 0)$. 
Thus $\Sigma$ admits no convex function. 
Note that if we are concerned only with the region $\rho > M/2$ 
(outside the trapped region) 
the restriction of the function 
\ben 
   \tilde{f} = \frac{1}{2} \left( 1 + \frac{M}{2 \rho} \right)^4  \rho^2 
\een 
is a strictly convex function.
However it ceases to be {\sl strictly} convex on the totally geodesic
submanifold at $\rho =M/2$. Restricted to $\rho=M/2$,
the function is a constant, and hence not (strictly) subharmonic.  
Because $\rho= M/2$ is closed and totally geodesic
it contains closed geodesics which are also closed geodesics
of the ambient manifold. In fact every such geodesic is a great circle
on the 2-sphere $\rho=M/2$. Again this shows that
the complete initial data set admits no strictly convex function.   

\vs 

From a  spacetime view point, $(\Sigma, h_{ij})$ in the example above 
is a time-symmetric hypersurface and the two-surface $S$ at $\rho = M/2$ 
is a marginally inner and outer trapped surface 
in the maximally extended Schwarzschild solution.
As an application to time symmetric initial data sets, 
consider $(\Sigma, h_{ij}, K_{ij})$ with the vanishing second fundamental 
form $K_{ij} = 0$. 
Then $S \subset \Sigma$ would have been a marginally inner and outer 
trapped surface. 
We can observe that if $\Sigma$ admits a convex function $\tilde{f}$, 
then no such apparent horizons can exist. 
This will be discussed further in the next section. 

\vs 
\noindent 
{\bf Level sets} \\ 
\noindent 
Level sets of convex functions have positive extrinsic curvature,
in other words, the second fundamental form is positive definite.  
We will discuss this further  in Section~\ref{sec:Levelsets}.

\subsection{Convex functions and Killing vector fields}

Killing vector fields are important in General Relativity
and it is interesting therefore that there are connections
with convex functions (see~[U] 
and references there-in). 

\begin{proposition} 
Let $\xi^i$ be a Killing vector field on $(\Sigma,h_{ij})$ and 
define 
\ben 
\tilde{f} := \frac{1}{2} \xi_i \xi^i \;.
\label{fnc:Killing}
\een
If the sectional curvature $R_{iljm} \xi^l \xi^m $ of $(\Sigma,h_{ij})$ 
is non positive,  then $\tilde{f}$ is (but
not necessarily strictly) convex. 
\end{proposition}

\bproof
From (\ref{fnc:Killing}) and the Killing equation $D_i \xi_j + D_j \xi_i =0$, 
we obtain 
\ben  
  D_i D_j \tilde{f} = D_i \xi^m D_j \xi_m - R_{iljm} \xi^l \xi^m \;. 
\een 
Hence $R_{iljm} \xi^l \xi^m \le 0$ yields $D_i D_j \tilde{f} \ge 0$.
\eproof 

\vs 

We shall give a few examples for which the function (\ref{fnc:Killing}) 
becomes a convex function. A simple example is a Killing vector field 
$\xi^i \partial_i = \partial_\tau$ on Euclidean Rindler space 
\ben 
   h_{ij}dx^idx^j = \alpha \chi^2 d\tau^2 + d\chi^2 \;, 
\een 
with a positive constant $\alpha$, for which we have 
\ben
   D_i D_j \tilde{f} = \alpha h_{ij} > \tilde{c} h_{ij} \;. 
\een 

Our next example is a Killing vector field, 
$
   \xi^i \partial_i = \partial_\tau \;, 
$
on an Euclidean anti-de-Sitter space or a hyperbolic space ${\Bbb H}^{n}$, 
\ben
  h_{ij}dx^idx^j = (1 + r^2) d\tau^2  + \frac{dr^2}{1 + r^2} 
                  + r^2 d\Omega^2_{(n-2)} \;. 
\een
The function ${1 \over 2} (1+r^2)$ and hence ${ 1\over 2} r^2$ are convex.

The function achieves its minimum at $r=0$, which is  a geodesic.
Moreover if one identifies $\tau$ (with any arbitrary period)
it becomes a closed  geodesic, along which $\tilde f$ is constant. 
Thus $\tilde f$ is not a {\sl strictly} convex function.  

Another example is Euclidean Schwarzschild spacetime 
\ben
  h_{ij}dx^idx^j = \left(1 - \frac{2M}{r} \right) d\tau^2  
                   + \left(1 - \frac{2M}{r} \right)^{-1} dr^2
                  + r^2 \Omega_{pq} dz^p dz^q \;.  
\een
If $r> 2M$, then the function (\ref{fnc:Killing}) with a Killing vector field 
$\xi^i \partial_i = \partial_\tau$ is convex.
However this function cannot be extended to the complete manifold
with $r\ge 2M$, $0\le \tau \le 8 \pi M$ as a {\sl strictly}
convex function because the two-dimensional fixed point set $r=2M$ 
is compact (indeed it is topologically and 
geometrically a 2-sphere) and totally geodesic. This means
it is totally geodesic and hence a closed minimal submanifold.
As we have seen this is incompatible with the existence of
a strictly convex function. 

\vs 

In general, a two-dimensional fixed point set of a Killing vector
field in a Riemannian 4-manifold is called 
a {\it bolt}~[GH]. 
We obtain the following  


\begin{proposition} 
A Riemannian 4-manifold admitting a strictly convex function can admit no 
Killing field with a closed bolt.
\end{proposition}

\section{Spacetime convex functions}  
\label{sect:spacetime} 

\subsection{Definitions and a canonical example}

We shall define a convex function on Lorentzian manifolds. 
We have two different definitions of convex function 
on an $(n+1)$-dimensional spacetime $(M,g_{\mu \nu})$ as follows. 

\begin{definition} {\bf Classical definition:} \\
A $C^\infty$ function $f : M \rightarrow {\Bbb R}$ is said to be 
{\em classically convex} if the Hessian is positive, i.e., 
\ben
(i) \quad  \nabla_\mu \nabla_\nu f > 0 \;, 
\een
where $\nabla_\mu$ is the Levi-Civita connection of 
$g_{\mu \nu}$. 
\end{definition}
\begin{definition}{\bf Spacetime definition:} \\
A smooth function $f: M \rightarrow {\Bbb R}$ is called 
a {\em spacetime convex function} 
if the Hessian $\nabla_\mu \nabla_\nu f$ has Lorentzian signature 
and satisfies the condition, 
\ben 
(ii) \quad  \nabla_\mu \nabla_\nu f \ge c g_{\mu \nu}
\een with a positive constant $c$, i.e., 
$
 V^\mu V^\nu \nabla_\mu \nabla_\nu f 
 \ge c g_{\mu \nu} V^\mu V^\nu \;, 
$ for ${}^\forall V^\mu \in TM$. 
\end{definition} 

\vs 

\begin{figure}[h] 
 \centerline{\epsfxsize = 5cm \epsfbox{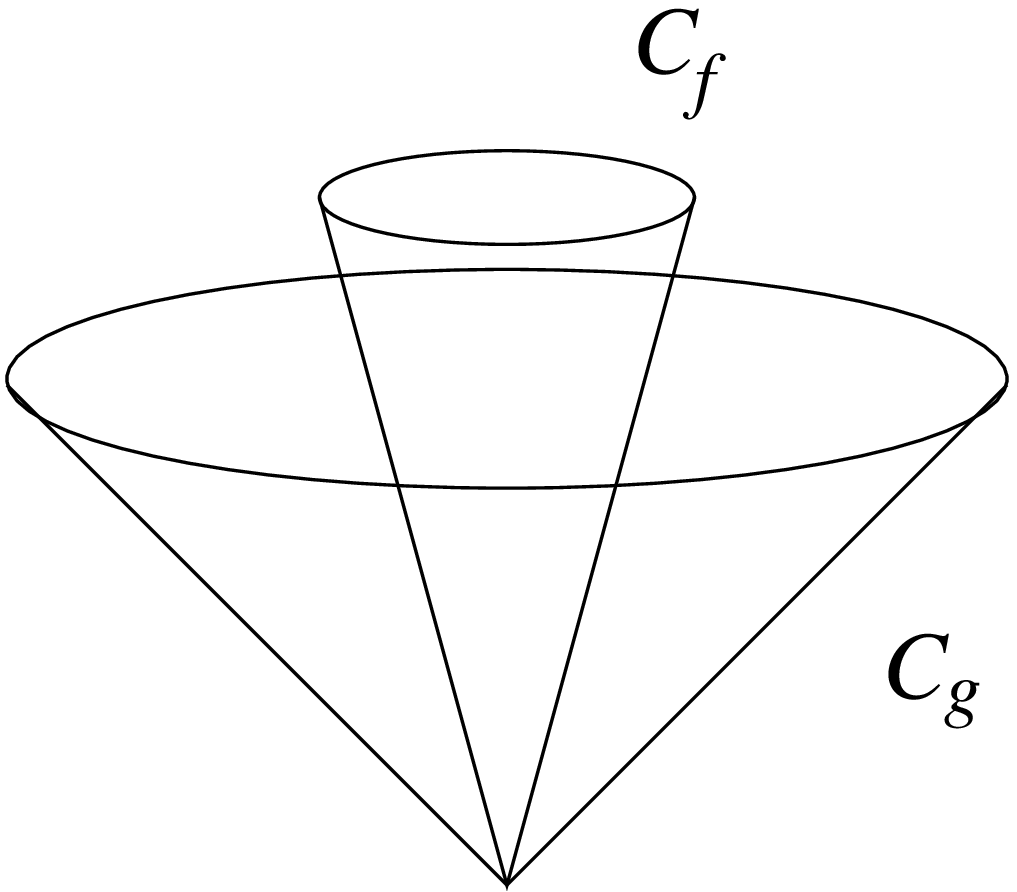}}
\vspace{3mm}
        \caption{The light cones ${\cal C}_f$ and ${\cal C}_g$.}
        \protect \label{fig:lightcone}
\end{figure}

The geometrical interpretation of the spacetime definition is that 
the forward light cone ${\cal C}_f$ defined by 
the Hessian~$ f_{\mu \nu} := \nabla_\mu \nabla_\nu f$ lies (strictly) 
inside the light cone ${\cal C}_g$ defined by the spacetime metric 
$g_{\mu \nu}$ as depicted in the figure~\ref{fig:lightcone}. 
This means, in some intuitive sense, that the condition (ii) prevents 
the ``collapse'' of the light cone defined by $g_{\mu \nu}$. 

\vs 

One of the simplest examples of a spacetime convex function 
is the canonical one, 
\ben
  f = \frac{1}{2} \left( x^i x^i - \alpha t^2 \right) \;, 
          \quad (t,x^i) \in {\Bbb E}^{n,1} \;, 
\label{convex:canonical}
\een
where $\alpha$ is a constant such that $0 < \alpha \le 1$.

The Hessian is given by 
\ben  
   f_{\mu \nu} = \eta_{\mu \nu} + (1-\alpha) t_\mu t_\nu  \;
\een 
where $\eta_{\mu \nu}$ is the Minkowski metric and 
$t_\mu := \partial_\mu t$, therefore $f_{\mu \nu}$ actually has 
Lorentzian signature and can define a light cone ${\cal C}_f$. 
In the particular case $\alpha=1$, $f_{\mu \nu} = \eta_{\mu \nu}$ 
and $c$ must be unit. 

Note that our conventions mean that, considered as a function of time, 
$f$ in (\ref{convex:canonical}) is what is conventionally called 
a ``concave function of time.'' However it is a conventional convex function 
of $x^i$. 

\vs 

In the canonical example~(\ref{convex:canonical}), 
~${\Bbb E}^{n,1} = ({\Bbb R}, \;-dt^2) \times {\Bbb E}^n$ and 
$f$ may be regarded as a direct sum of a concave function 
$ - \alpha t^2/2$ on $({\Bbb R},\;-dt^2)$ 
and a convex function $ x^i x^i/2$ on ${\Bbb E}^n$. 
In general, when a spacetime $(M,g)$ splits isometrically as a product 
$({\Bbb R} \times \Sigma, \; -dt^2 \oplus h)$, 
one can construct a spacetime convex function by summing concave functions 
of time $t\in {\Bbb R}$ and convex functions on $(\Sigma, h)$. 
However a direct sum of a convex cone ${C}_{\Bbb R}$ which consists 
of concave functions of time $t\in {\Bbb R}$ and a convex cone 
$C_{\Sigma}$ consisting of convex functions on $\Sigma$, i.e., 
\ben 
   {C}_{\Bbb R} \oplus C_{\Sigma} \;, 
\een 
is not a convex cone $C_M$ which consists of spacetime convex functions 
on $(M,g)$. This can be seen by considering the range of $\alpha$ 
in (\ref{convex:canonical}). 
(i) If we take $\alpha < 0$,  $f_{\mu \nu}$ becomes positive definite, 
hence $f$ is classically convex, 
and (ii) if $\alpha = 0$, $f_{\mu \nu}$ becomes positive semi-definite 
($t^\mu t^\nu f_{\mu \nu} = 0$), hence $f$ is not strictly classically 
convex, 
and (iii) if $\alpha > 1$, $f$ is not a convex function either 
in the classical or in the spacetime sense.

\subsection{Hypersurface orthogonal homothetic Killing vector field}  

\begin{proposition} 
If $f$ is a spacetime convex function and $\nabla^\mu f$ is a conformal 
Killing vector field, $\nabla^\mu f$ must be homothetic. 
\end{proposition} 

\bproof
Let $t^\mu$ and $s^\mu$ be any unit timelike and spacelike vector, 
respectively. Then, by definition of the spacetime convex function, 
we have $t^\mu t^\nu \nabla_\mu \nabla_\nu f \ge - c$, and 
$s^\mu s^\nu \nabla_\mu \nabla_\nu f \ge c$. 
On the other hand, when $\nabla_\mu f$ is a conformal Killing vector field, 
i.e., $\nabla_\mu \nabla_\nu f = \phi(x) g_{\mu \nu}$, 
it follows that $t^\mu t^\nu \nabla_\mu \nabla_\nu f = - \phi$, and 
$s^\mu s^\nu \nabla_\mu \nabla_\nu f = \phi$. 
Therefore $ \phi(x) = c$. 
\eproof  

\vs
It is easy to see that, if $f$ is classically convex, 
$\nabla_\mu f$ cannot be a Killing vector field. 

\vs 

Suppose that we have a hypersurface orthogonal homothetic vector field
$D$. An example is $ D =x^\mu {\partial \over \partial x^\mu} $ 
in flat Minkowski spacetime. In general there will be 
regions where $D$ is timelike and regions where it is spacelike, 
separated by a null hypersurface on which $D$ 
coincides with the null generators. In the region where it is timelike
one may introduce the parameter $\tau$ so that $D^\mu \nabla_\mu \tau =1$ 
and the metric takes the form
\ben
ds^2= -d\tau^2 + \tau^2 q_{ij}(x^k)dx^i dx^j \;, 
\label{metric:homothetic}
\een
where $q_{ij}$ is a positive definite $n$-metric.


\subsection{Spacetime convex functions and submanifolds} 

Suppose a spacetime convex function $f$ exists in $(M,g_{\mu \nu})$. 
Then, we deduce 
\begin{proposition}
 $(M,g_{\mu \nu})$ admits no closed spacelike geodesics. 
\end{proposition}

\bproof
The same argument as the Riemannian case.  
\eproof   
\vs 

It is a non trivial question whether or not $(M,g_{\mu \nu})$ admits 
a closed timelike geodesic. This will be discussed further 
in Section~\ref{sec:AdS}. 

\vs 

We also have 
\begin{proposition} 
Consider a spacetime $(M,g_{\mu \nu})$ and a closed spacelike 
surface $S \subset M$. Suppose that in a neighbourhood of $S$ 
the metric is written as 
\ben 
g_{\mu \nu}dx^\mu dx^\nu = \gamma_{ab} dy^a dy^b + k_{pq}dz^p dz^q \;, 
\een 
where $k_{pq}dz^p dz^q$ is the metric on $S$ and 
the components of the two-dimensional Lorentzian metric $\gamma_{ab}$ are 
independent of the coordinates $z^p$. 
Then, $S$ cannot be a closed marginally inner and outer trapped surface.  
\label{proposition:trappedsurface}
\end{proposition}

\bproof 
Let $l^\alpha$ and $n^\alpha $ satisfying 
$l^\alpha  l_\alpha  = n^\alpha  n_\alpha  = 0$, $l^\alpha n_\alpha  = -1$ 
be the two null normals of a closed spacelike surface $S$ 
so that $\gamma_{ab} = - l_a n_b - n_a l_b$. 
Then we have 
\ben
   \nabla^\mu  \nabla_\mu f = \gD{}^a \gD_a f + \kD{}^p \kD_p f 
          - ( l^a n^b + n^a l^b) \gD_a \log \sqrt{k} \: \gD_b f  \;, 
\label{Hessian:decomposition}
\een 
where $\gD_a$ denotes the Levi-Civita connection of $\gamma_{ab}$ 
and $k:= \det(k_{pq})$. 
Since $(M,g_{\mu \nu})$ admits a spacetime convex function $f$ 
and $\partial_p \gamma_{ab} = 0$, we have 
\ben
  \nabla_a  \nabla_b f = \gD_a \gD_b f \ge c \gamma_{ab} \;. 
\een 
Then from the partial trace 
\ben
   \nabla^a  \nabla_a f = 
                          \gD{}^2 f \ge 2c  \;, 
\een 
and (\ref{Hessian:decomposition}) we obtain 
\bena
   \kD{}^p \kD_p f 
     - ( l^a n^b + n^a l^b) \gD_a \log \sqrt{k} \: \gD_b f  
     = \nabla^\mu  \nabla_\mu f - \gD{}^a \gD_a f 
     \ge (n-1)c \;.
\label{ineq} 
\eena 
Since $\sqrt{k}$ gives the area of $S$, 
if $S$ is a closed marginally inner and outer trapped surface, 
it follows that 
\ben
l^a \gD_a \log \sqrt{k} = n^a \gD_a \log \sqrt{k} = 0 \;, 
\een
on $S$.  
Hence, it turns out from (\ref{ineq}) that $S$ admits a subharmonic 
function, i.e., $\kD{}^2 f \ge (n-1)c > 0$. 
This provides a contradiction. 
\eproof 

\vs 

This suggests that the existence of a spacetime convex function 
might be incompatible with a worm hole like structure. 

\vs 

In Subsection~\ref{subsec:submfds}, we pointed out that no marginally 
inner and outer trapped surface can exist on time symmetric 
initial hypersurface $\Sigma$, using a convex function $\tilde{f}$ 
on a Riemannian submanifold $\Sigma$ 
rather than a spacetime convex function $f$. 
We shall reconsider the previous situation. 
Suppose $(M,g_{\mu \nu})$ contains an embedded spacelike hypersurface 
$(\Sigma, h_{ij})$ with the second fundamental form $K_{ij}$. 
Then we have
\ben
 \nabla_i \nabla_j f = D_i D_j f + K_{ij} \frac{\partial f}{\partial t} \;, 
\een 
where $\partial f /\partial t := t^\mu \partial_\mu f$ with 
$t^\mu$ being a unit timelike vector, $t^\mu t_\mu = -1$, normal to $\Sigma$. 
Assuming that $(M,g_{\mu \nu})$ admits a spacetime convex function $f$, 
we deduce 
\begin{proposition} 
 If $\Sigma$ is totally geodesic, i.e., $\Sigma$ is a surface of time 
symmetry, then $(\Sigma, h_{ij})$ admits a convex function.  
\end{proposition}

\bproof 
This follows since $K_{ij} = 0$. 
\eproof 
\vs 

\noindent 
Moreover we have
\begin{proposition} 
 If $\Sigma$ is maximal, i.e., $h^{ij}K_{ij} = 0$, it admits 
 subharmonic function. 
\end{proposition}
\begin{corollary}
 $(\Sigma, h_{ij})$ cannot be closed. 
\label{coro:cannot be closed}
\end{corollary}
\begin{corollary}
 $(\Sigma, h_{ij})$ admits no closed geodesics nor minimal two surface 
in the case that $K_{ij}=0$. 
\label{corollary:minimalsurface}
\end{corollary}

As a simple example, consider the maximally extended Schwarzschild solution. 
In terms of the Kruskal coordinates 
$(U, V) = (-e^{-(t-r_*)/4M}, e^{(t+r_*)/4M})$, $r_*:= r + 2M \log(r/2M - 1)$, 
the metric is given by $k_{pq}dz^pdz^q = r^2 d\Omega^2_{(2)}$ and 
$\gamma_{ab} = - l_a n_b- n_a l_b$, where 

\ben 
l_a = - F \partial_a V, \quad n_a = -F \partial_a U, \quad
F^{2} := {16 M^3 e^{-r/2M} \over r } \;. 
\een 
Each $U+V = constant$ hypersurface $\Sigma$ is a time symmetric 
complete hypersurface which contains 
a marginally inner and outer trapped surface at $U=V=0$. 
Hence, this spacetime cannot admit a spacetime convex function. 
This also accords with the claim in Subsection~\ref{subsec:submfds} 
that the complete Schwarzschild initial data set admits no strictly 
convex function.

\section{Convex functions in cosmological spacetimes} 
\label{sec:cosmology}

Let us consider $(n+1)$-dimensional 
Friedmann-Lemaitre-Robertson-Walker~(FLRW) metric 
\ben
   ds^2 = \epsilon d\tau^2 + a(\tau)^2 q_{ij}dx^idx^j \;, 
\label{metric:FLRW}
\een 
where $q_{ij}dx^idx^j$ is a metric of $n$-dimensional space with 
constant curvature $K= \pm1,\,0$.  We begin by considering
functions which share the symmetry of the metric.

For an arbitrary function $f$ of $\tau$, we have 
\bena
\nabla_\tau \nabla_\tau f &=& \epsilon \ddot{f} g_{\tau \tau}\;, 
\label{Hess-cosmo:tt}  
\\
\nabla_\tau \nabla_i f &=& 0           \;, 
\label{Hess-cosmo:ti}
\\ 
\nabla_i\nabla_j f     &=& \epsilon \frac{\dot{a}}{a} \dot{f} g_{ij} \;, 
\label{Hess-cosmo:ij}
\eena  
where a dot denotes $\tau$ derivative. 
In what follows we are concerned in particular with the case that 
$\epsilon = -1$, $q_{ij}dx^idx^j$ is a Riemannian metric, 
and $\dot{a} > 0$: i.e., with an expanding universe.

Note the non-intuitive fact that what one usually thinks of as 
a ``convex function of time'', i.e., one satisfying 
\ben
   \dot{f} > constant > 0 \;, \;\; \ddot{f} > constant' > 0\;, 
\een 
is {\em never} a convex function in the classical sense because the RHS 
of (\ref{Hess-cosmo:ij}) is negative. Moreover $-f$ is {\em not} 
a concave function in the classical sense. 
However, the function $-f$ {\em is} a spacetime convex function, 
provided $\dot{a}$ is bounded below. This shows that if one wishes 
to use the idea of a convex function in cosmology one needs the 
spacetime definition.

We assume that the energy density $\rho$ and the pressure $p$ of 
matter contained in the universe obey $ p =w \rho $. 
Then from the Einstein equations $G_{\mu \nu} = \kappa^2 T_{\mu \nu}$ 
and the conservation law $\nabla_\nu T^\nu_\mu =0$ 
we have 
\ben
    \dot{a}^2 = \sigma^2 a^{2 - n(1 + w)} - K \;, 
\label{eq:Hubble}
\een 
where $\sigma^2 := 2\kappa^2 \rho_0/n(n-1)$ with $\rho_0$ being 
the value of $\rho$ at $\tau_0$. 
Naturally we require $\rho \ge 0$, so $\sigma^2 \ge 0 $. 
Note that the RHS is positive and is bounded below by some positive constant 
in the $K= -1$ case. 

Now let us consider a function 
\ben
   f = - \frac{1}{2} a^2 \;. 
\label{fnc:FLRWuniverse}
\een
If there exists a constant $c>0$ such that 
\bena
\nabla_\tau \nabla_\tau f &=& -\dot{a}^2 - a \ddot{a} \ge -c \;, 
\label{Hessian:tt} 
\\
 \{ \nabla_i \nabla_j f \} /g_{ij} &=& \dot{a}^2  \ge c \;. 
\label{Hessian:ij} 
\eena
then the function $f$ is convex in the spacetime sense.

\vs \noindent 
{\bf (i) $\ddot{a} = 0$: uniformly expanding universe.} \\ 
From (\ref{Hessian:tt}) and (\ref{Hessian:ij}) we find that 
$\dot{a}^2 = c$ and the function (\ref{fnc:FLRWuniverse}) 
is a spacetime convex function. 
In this case $\nabla_\mu \nabla_\nu f = c g_{\mu \nu}$, i.e., 
$\nabla_\mu f$ is a homothetic vector field, and 
the metric is the special case of~(\ref{metric:homothetic}).

From (\ref{eq:Hubble}), it turns out that 
such a convex function is allowed on a vacuum open ($\sigma^2 = 0,\;K=-1$) 
FLRW universe, i.e., the Milne universe, and 
on flat and closed ($K=0,\;+1$) FLRW universes with the matter satisfying 
\ben 
   w = \frac{2-n}{n} \;, \quad \sigma^2 = c + K  \;, 
\een 
which is realized as, for example, a cosmic string dominated universe 
in the 4-dimensional case.

\vs \noindent 
{\bf (ii) $\ddot{a} >0$: accelerately expanding universe.} \\ 
In this case 
(\ref{Hessian:tt}) and (\ref{Hessian:ij}) cannot hold simultaneously 
since $a \ddot{a}>0$. Hence $f$ cannot be a spacetime convex function.
This implies that inflationary universes do not admit 
a convex function of the form~(\ref{fnc:FLRWuniverse}).

\vs \noindent 
{\bf (iii) $\ddot{a} <0$: decelerately expanding universe.} \\ 
From (\ref{Hessian:tt}) and (\ref{Hessian:ij}), 
if we can find a constant $c>0$ which satisfies the condition
\ben 
   0 \le \dot{a}^2 - c \le - a \ddot{a} \;, \quad 
   - a \ddot{a} < \dot{a}^2 \;, 
\label{condi:deceleration case}
\een 
the function (\ref{fnc:FLRWuniverse}) becomes a spacetime convex function. 
Note that when the first condition holds but the second does not, 
the Hessian $\nabla_\mu \nabla_\nu f$ becomes positive semi-definite 
and does not have Lorentzian signature. 
Note also that since (\ref{eq:Hubble}) and its differentiation give 
\ben  
   a \ddot{a} = \frac{2-n(1+w)}{2} \sigma^2 a^{2-n(1+w)} \;, 
\een 
the assumption of deceleration $\ddot{a}<0$ requires that 
\ben
    w > \frac{2-n}{n} \;, \quad \sigma^2 > 0 \;, 
\een 
which corresponds to the timelike convergence condition, 
\ben
  R_{\tau \tau} 
  = \kappa^2 \left( T_{\tau \tau} + \frac{1}{n-1}T^\mu_\mu \right) > 0 \;. 
\een 

In terms of $w$ and $\sigma^2$, 
the condition~(\ref{condi:deceleration case}) is written as 
\ben
 0 \le \frac{\sigma^2}{a^{n(1+w)-2}} - K - c   
   \le \frac{n(1+w)-2}{2} \frac{\sigma^2}{a^{n(1+w)-2}} 
   < \frac{\sigma^2}{a^{n(1+w)-2}} - K   \;. 
\label{condi:deceleration w}
\een 
Considering a suitable open interval $I \subset {\Bbb R}$, 
we can find a constant $c$ satisfying (\ref{condi:deceleration w}) 
for $\tau \in I$. Then $f$ becomes a spacetime convex function 
on $I\times \Sigma \subset M$. 

For example, when a radiation dominated open FLRW 
($n=3,\, w = 1/3, \, K = -1$) universe is considered, 
the condition (\ref{condi:deceleration w}) is reduced to  
\ben
  1 \le c \le 1 + \frac{\sigma^2}{a^2} \;. 
\een 
Then we can choose $c =1$ for $I=(0,\infty)$ so that $f$ becomes a spacetime 
convex function on $I \times \Sigma$, which covers the whole region of $M$. 
This can be, of course, verified directly by examining the solution 
$a = \sigma \{ (1+ \sigma^{-1}\tau)^2 - 1 \}^{1/2}$ of (\ref{eq:Hubble}).

In general, $I \times \Sigma$ cannot be extended to cover the whole universe 
while keeping $f$ a spacetime convex function. 
For example, in a closed ($K=1$) FLRW case the interval $I$ should not contain 
the moment of maximal expansion, where $\dot{a}$ vanishes. 
Indeed the timeslice of maximal expansion is a closed maximal hypersurface 
and according to Corollary~\ref{coro:cannot be closed} 
such an $I\times \Sigma$ admits no spacetime convex functions. 

In the following sections we shall discuss the issue of the global existence 
of convex functions, examining de-Sitter and anti-de-Sitter spacetime.

\section{Convex functions and de-Sitter spacetime}    
\label{sec:deS}

In this section we shall consider convex functions in de-Sitter
spacetime, which is a maximally symmetric spacetime 
with positive constant curvature.

\vs \noindent
{\bf(i): Cosmological charts} \\ 
We can introduce the three types of cosmological charts for which 
the metric takes the form of (\ref{metric:FLRW}) with 
$\epsilon = -1$ and $q_{ij}dx^idx^j$ being a Riemannian metric. 
The scale factors are given by 
\ben 
   a(\tau) =  \cosh \tau \;, \quad e^\tau \;, \quad \sinh \tau \;, 
\een 
for $K= 1,\,0,\,-1$, respectively. 
Note that the closed ($K=1$) FLRW chart covers the whole 
de-Sitter spacetime but the $K=0$ and $K=-1$ FLRW charts do not. 

In these cosmological charts, the function $f$ of 
the form~(\ref{fnc:FLRWuniverse}) cannot be a spacetime convex function. 
This agrees with the argument of the previous section 
since $\ddot{a} >0$ in the expanding region where $\dot{a}>0$.

\vs \noindent  
{\bf (ii): Static chart} \\  
We can choose a coordinate system so that the metric has the static form, 
\ben
  ds^2 = - (1 - r^2) dt^2 
       + \frac{dr^2}{(1 - r^2)} + r^2 \Omega^{(n-1)}_{pq}dz^pdz^q \;. 
\een   
The Hessian of the function 
\ben 
    f = \frac{1}{2}r^2 
\een
is given by 
\ben
  \nabla_t \nabla_t f = (1-r^2)r^2 \;, \quad 
  \nabla_r \nabla_r f = (1- 2r^2)g_{rr} \;, \quad 
  \nabla_p \nabla_q f = (1- r^2)g_{pq} \;. 
\een
Since $\nabla_t \nabla_t f$ is positive in the region $0 < r < 1$, 
where $g_{tt}<0$, $f$ cannot be a spacetime convex function, though 
it can become classically convex in the region $0 < r < 1/2$.

\vs \noindent
{\bf (iii): Rindler chart} \\ 
The de-Sitter metric can take the following form 
\ben
ds^2 = d\tau^2 + \sin^2 \tau
      \left\{ -d\psi^2 + \cosh^2 \psi d\Omega_{(n-1)}^2 \right\}\;.
\een 
This corresponds to the metric (\ref{metric:FLRW}) with 
$\epsilon = +1$, $q_{ij}dx^idx^j$ being a Lorentzian metric of $K=+1$,
and $a(\tau) = \sin \tau$. 
Let us consider the function 
\ben
   f = \frac{1}{2}a^2 \;. 
\een
Then from (\ref{Hess-cosmo:tt}) and (\ref{Hess-cosmo:ij}), 
we can see that $f$ would be a spacetime convex function 
if there was a constant $c>0$ such that 
$c \le \dot{a}^2 + a \ddot{a} , \, 
\dot{a}^2 \le c \le \dot{a}^2$. 
There is no such $c$ since $a \ddot{a}<0$, hence $f$ is not spacetime convex.

\section{Convex functions in anti-de-Sitter spacetime} 
\label{sec:AdS}  

We shall now examine the existence of convex function on
anti-de-Sitter spacetime, 
which is maximally symmetric spacetime with negative constant curvature.

The metric of $(n+1)$-dimensional anti-de-Sitter spacetime may be given 
in the cosmological (open FLRW) form, 
\ben  
 ds^2 = - d\tau^2 + \sin^2 \tau 
         \left( d\xi^2 + \sinh^2\xi d \Omega_{(n-1)}^2 \right) \;, 
\label{chart:AdS-Open}
\een 
where $ d \Omega_{(n-1)}^2$ is the metric of a unit $(n-1)$-sphere. 
This is the case that $\epsilon = -1$, 
$q_{ij}dx^idx^j$ is the Riemannian metric with constant curvature $K=-1$, 
and $a(\tau) = \sin \tau$ in (\ref{metric:FLRW}).  
Since $\dot{a}>0,\; \ddot{a}<0$ for $\tau \in (0,\pi/2)$, 
according to the argument in Section~\ref{sec:cosmology}, 
the function 
\ben
 f = - {1 \over 2} a^2 = - {1 \over 2} \sin^2 \tau 
\label{fnc:convex-in-AdS}
\een  
can be spacetime convex. 
Indeed, the Hessian is given by 
\ben
\nabla_\tau \nabla_\tau f = - \cos 2 \tau\;, \;\;
\nabla_\tau \nabla_i f = 0           \;, \;\;
\nabla_i\nabla_j f     = \cos^2 \tau g_{ij} \;, 
\een
hence (\ref{fnc:convex-in-AdS}) is spacetime convex at least locally. 
If the time interval $I$ contains a moment $\tau = \pi/2$, then 
(\ref{fnc:convex-in-AdS}) is no longer strictly convex 
in such a $I \times \Sigma$. 

\vs 

Anti-de-Sitter spacetime admits closed timelike geodesics. 
Therefore the existence of a locally spacetime convex function 
does not exclude closed timelike geodesics.

\begin{figure}[h] 
 \centerline{\epsfxsize = 2.5cm \epsfbox{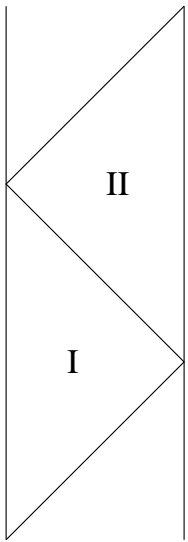}}
\vspace{3mm}
        \caption{The Penrose diagram of the universal covering space of 
        anti-de-Sitter spacetime. A single and double vertical lines 
        denote a centre of the spherical symmetry and the infinity, 
        respectively.}  
        \protect 
        \label{fig:AdS} 
\end{figure}

The cosmological chart (\ref{chart:AdS-Open}) does not cover the whole 
spacetime; it covers the region~I as illustrated in the figure~\ref{fig:AdS} 
and the boundary is a Cauchy horizon for $\tau = constant$ hypersurfaces. 
To see whether or not the locally spacetime convex function $f$ 
discussed above is extendible beyond the horizon while maintaining 
its spacetime convexity, let us consider the embedding of $(n+1)$-dimensional 
anti-de-Sitter spacetime in ${\Bbb E}^{n,2}$ as a hyperboloid: 
\ben 
  - (X^0)^2 - (X^1)^2 + (X^1)^2 + \cdots + (X^{n+1})^2 = - 1 \;, 
\een 
where $X^a$ are the Cartesian coordinates of ${\Bbb E}^{n,2}$. 
The open FLRW chart (\ref{chart:AdS-Open}) is then introduced by  
\bena
    X^0 = \cos \tau \;, \;\; 
    X^i = \sin \tau z^i \;, \quad (i = 1, ..., n+1) \;, 
\eena 
where $z^i \in {\Bbb H}^n$, i.e., 
$ - (z^1)^2 + (z^2)^2 + \cdots + (z^{n+1})^2 = -1$. 
One can observe that $ |X^0| < 1$ in region~I, $ X^0 = 1$ at the horizon, 
and $ 1< |X^0| $ beyond the horizon (region~II). 

In terms of the embedding coordinates, 
the locally convex function (\ref{fnc:convex-in-AdS}) is expressed as 
\ben 
 f = - {1 \over 2} \left\{1- (X^0)^2 \right\} \;. 
\label{fnc:global-extension}
\een  
This expression is not restricted to $|X^0| < 1$ and 
hence can be regarded as an extension of (\ref{fnc:convex-in-AdS}) 
beyond the horizon. 

Now, let us consider a function $\phi = X^0$. We see that 
\ben
   \xi^\mu := \nabla^\mu \phi \;, 
\een 
is a conformal Killing vector field, i.e., 
$ \nabla_\mu \xi_\nu = \nabla_\mu \nabla_\nu \phi = \sigma g_{\mu \nu}$, 
and, moreover, $\sigma = \phi$. 
Indeed, in the region~I where $\phi = \cos \tau$, we can directly verify 
the equation 
\ben 
\nabla_\mu \nabla_\nu \phi = \phi g_{\mu \nu} \;.
\een 
Since this expression is covariant and $\phi$ is globally defined, 
this equation holds globally. 
Note that $\xi^\mu$ becomes timelike in region~I, null at the horizon, 
and spacelike in region~II, since its norm is given by 
$\xi^\mu \xi_\mu = -1 + \phi^2$. 

Now, rewriting (\ref{fnc:global-extension}) as $ f = - (1 - \phi^2)/2 $, 
we obtain 
\ben
\nabla_\mu \nabla_\nu f = \phi \nabla_\mu \nabla_\nu \phi 
                        + \nabla_\mu \phi \nabla_\nu \phi 
                        = \phi^2 g_{\mu \nu} + \xi_\mu \xi_\nu \;. 
\label{Hessian:conformal}
\een
This is the global expression for the Hessian of $f$. 
In region~II, $\xi^\mu$ becomes spacelike and there is a unit timelike 
vector field $t^\mu$ which is orthogonal to $\xi^\mu$, i.e., 
$t^\mu \xi_\mu =0$. 
Then, 
contraction of (\ref{Hessian:conformal}) with $t^\mu$ gives 
$ t^\mu t^\nu\nabla_\mu \nabla_\nu f = - \phi^2 $. 
Since $\phi^2 = (X^0)^2$ is not bounded above in region~II, 
for any fixed ${c}>0$, 
$t^\mu t^\nu\nabla_\mu \nabla_\nu f \ge {c} g_{\mu \nu}t^\mu t^\nu$ 
fails in the region where $\phi^2 > {c}$. 
Thus, the function (\ref{fnc:global-extension}) cannot be globally 
spacetime convex.

\section{Convex functions and black hole spacetimes} 
\label{sec:BH}  

Let us consider $(n+1)$-dimensional static metric; 
\ben
ds^2 = \epsilon {\triangle} dt^2 
       + \frac{dr^2}{\triangle} + R^2(r) d \Omega^2_{(n-1)} \;, 
\label{metric:staticBH}
\een
where $\epsilon = \pm 1$, ${\triangle}$ is a function of $r$, 
and $d\Omega^2_{(n-1)} = \Omega_{pq}dz^p dz^q$ is 
the metric of ${(n-1)}$-dimensional space with unit constant curvature $K$.

\vs 
For an arbitrary function $f=f( t,r)$, we have 
\bena  
  \nabla_t \nabla_t f &=& \left( \frac{\epsilon}{\triangle} \ddot{f} 
 + \frac{\triangle'}{2} f' \right) g_{t t} \; , 
\label{Hessian:bh:tt}
\\
  \nabla_r \nabla_r f    &=& \left( 
                              \triangle f'' + \frac{\triangle'}{2} f' 
                             \right) g_{r r} \;, 
\label{Hessian:bh:rr}
\\
  \nabla_p \nabla_q f    &=& \triangle \frac{R' }{R} f' g_{pq} \;, 
\label{Hessian:bh:pq}
\eena 
where the {\it dot} and the {\it prime} denote $t$ and $r$ derivatives, 
respectively.

Consider a function 
\ben
    f = \frac{1}{2} \left( R^2  - \alpha t^2 \right) \;, 
\label{spacetime-convexfnc:BH} 
\een 
outside the event horizon, where $\triangle > 0$.  
This function asymptotically approaches 
the canonical one~(\ref{convex:canonical}).  
We see from (\ref{Hessian:bh:tt}), (\ref{Hessian:bh:rr}), and 
(\ref{Hessian:bh:pq}) that $f$ can be a spacetime convex function 
if there exists a constant $c>0$ satisfying 
\ben
     0<\frac{\alpha}{\triangle} + \frac{\triangle'}{2}RR' \le c \;, \quad 
     \triangle(R'^2 + RR'') + \frac{\triangle'}{2}RR' \ge c \;, \quad 
     \triangle R'^2 \ge c \;. 
\label{condition:bh:convex}
\een 


Let us examine the function~(\ref{spacetime-convexfnc:BH}) 
on the $(n+1)$-dimensional ($n \ge 3$) Schwarzschild spacetime, 
whose metric is given by 
\bena
  \epsilon = -1 \;, \quad 
  \triangle = 1 - \left( \frac{r_g}{r}\right)^{n-2} \;, \quad 
  R = r \;, 
\label{metric:Schwarzschild}
\eena 
where $r_g$ is the radius of the event horizon. 
Since $\triangle$ and ${\alpha}/{\triangle} +{\triangle'r}/{2}$ 
are monotonically increasing and decreasing function, respectively, 
one can observe that, for $0 \le \alpha < 1$, 
there exists a radius $r_* \,(> r_g)$:  
\ben 
r_*^{n-2} = \frac{n+2+ \sqrt{(n+2)^2 - 8n (1- \alpha)} }{4(1-\alpha)} r_g^{n-2}
\;, 
\een  
for which 
\ben 
\triangle(r_*) = \frac{\alpha}{\triangle (r_*)} 
                + \frac{\triangle'(r_*)}{2} r_*  \;. 
\een 
Then, if one chooses  
\ben
    c = \triangle(r_*) \;,      
\een
condition~(\ref{condition:bh:convex}) holds in the region 
$r_* \le r < \infty$, hence $f$ becomes a spacetime convex function there. 

Note that, in the black hole spacetime, the Hessian 
of (\ref{spacetime-convexfnc:BH}) has Lorentzian signature 
even for $\alpha=0$, while in the strictly flat spacetime, 
the Hessian becomes positive semi-definite for $\alpha =0$.

One can also see that, 
for $\alpha<0$, the condition (\ref{condition:bh:convex}) cannot hold 
in the asymptotic region, and for $1 \le \alpha $, 
(\ref{condition:bh:convex}) holds nowhere.

\vs 

If one takes $\triangle$ in (\ref{metric:Schwarzschild}) as 
\ben
   \triangle = 1 + \left( \frac{r_g}{r}\right)^{n-2} \;, 
\een 
then the metric (\ref{metric:staticBH}) describes the negative mass 
Schwarzschild spacetime, which has a timelike singularity and 
no horizon.  

One can observe that 
the function $\{ \nabla_t \nabla_t f\}/g_{tt}$ is a 
monotonically increasing function with the range $(- \infty, \alpha)$,  
and $\{\nabla_p \nabla_q f\}/g_{pq}$ monotonically decreasing
function with the range $(1,\infty)$.  
On the other hand, the function $\{\nabla_r \nabla_r f\}/g_{rr} $
is a decreasing function with the range $(1, \infty)$ in $n=3$ case 
and constant $=1$ in $n=4$ case, 
while it becomes monotonically increasing with the range 
$(-\infty, 1)$ in the case $n \ge 5$.

Therefore, in the cases $n= 3,4$, if $\alpha \le c \le 1$, then 
the function~(\ref{spacetime-convexfnc:BH}) with $(0< \alpha \le 1)$ 
is spacetime convex in the region $\hat{r}_* < r <\infty$, 
where $\hat{r}_*$ is given by 
$\{\nabla_t \nabla_t f\}/g_{tt} (\hat{r}_*) = 0 $.   
In the case $n \ge 5$, if one chooses $\alpha \le c <1$, then 
the function (\ref{spacetime-convexfnc:BH}) with $(0< \alpha < 1)$ 
is a spacetime convex function in the region 
$\max\{\hat{r}_*,\hat{r}_\star \} < r < \infty$, 
where $\hat{r}_\star$ is given by 
$\{\nabla_r \nabla_r f\}/g_{rr} (\hat{r}_\star) = c$. 
Thus we were not able to find a spacetime convex function 
through the region $r>0$ of the negative mass Schwarzschild 
solution.

\section{Level sets and foliations} 
\label{sec:Levelsets} 

Locally the level sets $\Sigma_c=\{x \in M| f(x) ={\rm constant}=c\}$ of 
a function $f$ have a unit normal given by 
\ben
n_\mu = - {\nabla _\mu f \over 
\sqrt{\epsilon \nabla _\nu f \nabla ^\nu f} 
}, 
\een
where $\epsilon =-1$ if the normal is timelike and 
$\epsilon =1$ if it is spacelike. 
Given two vectors $X^\mu$ and $Y^\nu$ 
tangent to $\Sigma_c$ so that 
$X^\mu \partial _\mu f=0=Y^\mu \partial _\mu f$ one may evaluate
 $K_{\mu \nu}X^\mu Y^\mu$ in terms of the Hessian of $f$:
\ben
K_{\mu \nu}X^\mu Y^\mu= X^\mu Y^\nu {\nabla _\mu \nabla _\nu 
f \over \sqrt{\epsilon \nabla _\nu f
\nabla ^\nu f}}.
\een
             
Thus, in the Riemannian case (for which the metric $g_{\mu \nu}$ on $M$ 
is positive definite and $\epsilon= 1$), 
a strictly convex function has a positive definite 
second fundamental form. 
Of course there is a convention here about the choice
of direction of the normal. We have chosen $f$ to decrease along
$n^\mu$. The converse is not necessarily true,
since $f(x)$ and $g(f(x))$ have the same level sets, where $g$ is 
a monotonic function of ${\Bbb R}$. Using this gauge freedom 
we may easily change the signature  of the Hessian. 
However, given a hypersurface $\Sigma_0$ 
with positive definite second fundamental form,
we may, locally,  also use this gauge freedom to find 
a convex function  whose level $f=0$  coincides with $\Sigma_0$. 
If we have a foliation (often called a ``slicing"   
by relativists) by hypersurfaces with positive definite fundamental
form, we may locally represent the leaves as the levels sets of a 
convex function. 
   
Similar remarks apply for  Lorentzian metrics.
The case of greatest interest is when the level sets have a timelike
normal.  For a classical strictly convex function, the second fundamental 
form will be positive definite and the hypersurface orthogonal timelike
congruence given by the normals $n^\mu$ is an expanding one.
For a spacetime convex function, our conventions also imply that
the second fundamental form of a spacelike level set 
is positive definite. This can be illustrated by the canonical
example (\ref{convex:canonical}) with $\alpha =1$ in flat spacetime. 
The spacelike level sets foliate the interior of the future (or past) 
light cone. Each leaf is isometric to hyperbolic space. 
The expansion is homogeneous and isotropic, and
if we introduce coordinates adapted to the foliation we
obtain the flat metric in $K=-1$ FLRW form (\ref{metric:FLRW}) 
with scale factor $a(\tau)=\tau$. This is often called the Milne model.
The convex function underlying this FLRW coordinate system is 
globally well defined but the coordinate system based on it 
breaks down at the origin of Minkowski spacetime.

From what has been said it is clear that there is 
a close relation between the existence of convex functions
and the existence of foliations with positive definite second 
fundamental form.
A special case, often encountered in practice,
are totally umbilic foliations. These have 
$K_{ij}= { 1\over n}g_{ij} {\rm Tr} K$. 
In a 4-dimensional Einstein space for which 
$R_{\mu \nu}=\Lambda g_{\mu \nu}$ the Gauss-Codazzi equations imply 
that 3-dimensional umbilic hypersurface $\Sigma_c$ is 
a constant curvature space with a sectional curvature $C$ 
and the trace of $K_{ij}$ is given by 
${\rm Tr} K =  3 \sqrt {{\Lambda \over 3} - C}$. 
The sectional curvature $C$ could depend only upon ``time function'' $f$ 
and thus totally umbilic hypersurfaces are a special case of 
``trace $K$ = constant" 
hypersurfaces which are used in numerical relativity.
%
One of the simplest cases is given by considering in 
Minkowski space, in which trace $K$ constant, often refered to as 
``constant mean curvature'' hypersurfaces are hyperbolic 
and convex~(see e.g. [T] and references therein). 
The Milne slicing mentioned above is such an obvious example. 
%
In the case of de-Sitter or anti-de-Sitter  spacetimes,
which may be represented by quadrics in five dimensions,
one may obtain totally umbilic hypersurfaces  
by intersecting the quadric with a hyperplane.
Many standard coordinate systems for de-Sitter and anti-de-Sitter
spacetime are obtained in this way. The singularities of these
coordinate systems may sometimes be seen 
in terms of the non-existence of convex functions.

\section{Barriers and Convex functions}
\label{sec:barries}

In the theory of maximal hypersurfaces and more generally
hypersurfaces of constant mean curvature (`` $ {\rm Tr} K =$ constant''
hypersurface) an important role is played by the idea of a
``barriers.'' The basic idea is that, if two spacelike hypersurfaces 
$\Sigma_1$ and $\Sigma_2$ touch, then the one in the future, 
$\Sigma_2$ say, can have no smaller a mean curvature than the 
$\Sigma_1$, the hypersurface in the past, i.e., 
${\rm Tr} K_1 \le {\rm Tr} K_2$. 
This is illustrated in the accompanying figure~\ref{fig:2surfaces}: 
\begin{figure}[h] 
 \centerline{\epsfxsize = 6cm \epsfbox{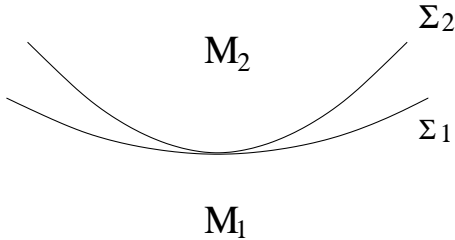}} 
\vspace{3mm}
        \caption{Spacelike hypersurfaces, $\Sigma_1$ and $\Sigma_2$. }  
        \protect 
        \label{fig:2surfaces} 
\end{figure}

More formally, we have, using the Maximum Principle, the following 
theorem: 

\begin{theorem}{\bf Eschenburg~[E]:}  
Let $M_1$ and $M_2$ be disjoint open domains with spacelike connected 
$C^2$-boundaries having a point in common. If the mean curvatures 
${\rm Tr} K_1$ of $\partial M_1$ and ${\rm Tr} K_2$ of $\partial M_2$ satisfy 
\ben 
   {\rm Tr} K_1 \le -a, \quad {\rm Tr} K_2 \le a \;,  
\een  
for some real number $a$, then $\partial M_1 = \partial M_2$, and 
${\rm Tr} K_2 = - {\rm Tr} K_1 = a$. 
\end{theorem}
 
\begin{figure}[h] 
 \centerline{\epsfxsize = 7cm \epsfbox{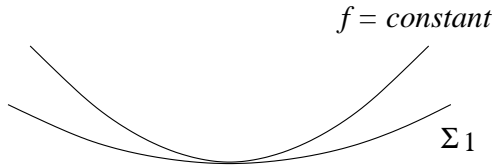}} 
\vspace{3mm}
        \caption{One of the level sets of $f$ and 
         a maximal hypersurface $\Sigma_1$. }  
        \protect 
        \label{fig:Concavesurface} 
\end{figure}

As an example, let us suppose that $\Sigma_1$ is maximal, i.e., 
${\rm Tr} K_1 =0$, where $K_1$ is the second fundamental form of 
$\Sigma_1$. Now consider a function $f$ whose level sets lie in 
the future of $\Sigma_1$ and one of which touches $\Sigma_1$ as in 
the figure~\ref{fig:Concavesurface}.   

The second fundamental form $K_f$ of the level set $f=$constant must have 
positive trace, ${\rm Tr} K_f > 0$. 
In particular, ${\rm Tr} K_f <0$ is excluded. 
Now if $f$ is a spacetime {\it concave} function, i.e., 
\ben
   \nabla_\mu \nabla_\nu f < -c g_{\mu \nu} \;, \quad  
\een 
for $c>0$, 
we have ${\rm Tr} K_f<0$. 
It follows that the level sets of a concave function cannot penetrate
the ``barrier'' produced by the maximal hypersurface $\Sigma_1$. 

As an application of this idea, consider the interior region of the 
Schwarzschild solution. 
The hypersurface $r=$ constant has 
\ben
   {\rm Tr} K(r) = - {2 \over r}\left({2M \over r} -1 \right)^{-1/2} 
                          \left(1 - {3M \over 2r} \right) \;. 
\een 
The sign of ${\rm Tr} K(r)$ is determined by the fact that regarding $r$ as
a time coordinate, $r$ decrease as time increases. Thus for $ r >
3M/2$, ${\rm Tr} K$ is negative but for $r <3M/2$, ${\rm Tr} K$ is positive. 
If $r = 3M/2$, we have ${\rm Tr} K=0$, that is, $r = 3M/2$ is a maximal
hypersurface. 

\begin{figure}[h] 
 \centerline{\epsfxsize = 9cm \epsfbox{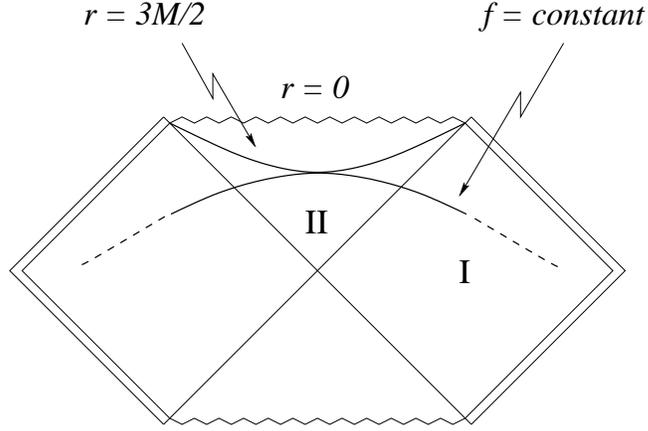}} 
\vspace{3mm}
        \caption{Penrose diagram of the Schwarzschild black hole.  
                 Level sets of a concave function
        $f$ must be in the past of the barrier hypersurface $r=3M/2$.}  
        \protect 
        \label{fig:Barriers} 
\end{figure}

Consider now attempting to foliate the interior region II 
by the level sets of a concave function. Every level set must touch 
an $r =$ constant hypersurface at some point. 
If $r = 3M/2$ this can, by the theorem, only happen if the level set 
lies in the past of the hypersurface $r = 3M/2$ as depicted in 
the figure~\ref{fig:Barriers}.  
Evidently therefore a foliation by level sets of a spacetime 
{\it concave} function can never penetrate the barrier at $r = 3M/2$. 
In particular, such a foliation can never extend to the singularity at
$r=0$. 

These results are relevant to work in numerical relativity. One
typically sets up a coordinate system in which the constant time
surfaces are of constant mean curvature, i.e., constant trace $K$. 
It follows from our results that, if the constant is {\sl negative}, 
then, the coordinate system can never penetrate the region $r<3M/2$. 
In fact it could never penetrate any maximal hypersurface.

\vs 

Now we shall illustrate the remarks mentioned above with an example of
a constant mean curvature foliation in a black hole spacetime. 
We are concerned with a spacelike hypersurface $\Sigma \subset M$ 
whose mean curvature is constant $\lambda$ (or zero). 
A convenient way to find such hypersurfaces in a given spacetime 
is to use the variational principle. 
In this approach, the behaviours of constant mean curvature slices 
in the Schwarzschild black hole have been studied 
by Brill {\it et.al}~[BCI]. 

Let us start from an action 
\ben 
 S = \int_\Sigma d^nx \sqrt{h} + \lambda \int_M d^{n+1}x \sqrt{-g}\;, 
\label{action:NG-constraint}
\een 
where $h$ and $g$ denote the determinant of the metric $h_{ij}$ on
$\Sigma$ and that of a spacetime metric $g_{\mu \nu}$, respectively. 
In other words, the first term is the volume of $\Sigma$ and the second
integral represents a spacetime volume enclosed by $\Sigma$ together with 
any fixed hypersurface. The variation of this action gives 
\ben
   n_\mu h^{ij} ( D_i \partial_jx^\mu 
     + \Gamma^\mu{}_{\nu \sigma} \partial_ix^\nu \partial_jx^\sigma)
         = \lambda \;,  
\label{eq:differential}
\een 
where $n^\mu$ is a unit normal vector to $\Sigma$ and 
$\Gamma^\mu{}_{\nu \lambda}$ is the Christoffel symbol of $g_{\mu \nu}$. 
Using the Gauss-Weingarten equation, one can see that 
(\ref{eq:differential}) is equivalent to 
\ben
   {\rm Tr} K = \lambda \;. 
\label{eq:K=lambda}
\een 
Thus, the solutions of the differential equation~(\ref{eq:differential}) 
describe constant mean curvature hypersurfaces $\Sigma$.

Consider now an $(n+1)$-dimensional black hole metric~(\ref{metric:staticBH}), 
\ben
   g_{\mu \nu} dx^\mu dx^\nu = - {\triangle} dt^2 + \frac{dr^2}{\triangle} 
          + R^2(r) \Omega_{pq}dz^pdz^q \;, 
\label{metric:blackhole}
\een 
and look for $\Sigma$, 
which we assume to be spherically symmetric, for simplicity. 
As we will see, the first integral of (\ref{eq:differential}) 
is easily obtained because the metric (\ref{metric:blackhole}) 
has a Killing symmetry.

Let us choose the coordinates $x^i$ on $\Sigma$ as $(t, z^p )$, 
then $\Sigma$ is described by $r =r(t)$ and the induced metric is written as 
\bena 
  h_{ij}dx^idx^j = \left(
                         -\triangle + {\triangle}^{-1} {\dot{r}^2}
                   \right)dt^2 
                         + R^2(r) d\Omega_{(n-1)}^2 \;, 
\eena  
where the {\it dot} denotes the derivative by $t$. 
Then, the action~(\ref{action:NG-constraint}) takes the form, 
\ben
 S = \int dt L 
   = \int dt \left\{ 
                R^{n-1} \sqrt{-\triangle + {\triangle}^{-1}{\dot{r}^2} }
                - \lambda F
             \right\} \;, \quad 
             F(r) := \int^r d\bar{r} R^{n-1}(\bar{r}) \;, 
\een 
where the integration of angular part is omitted. 
The coordinate $t$ is cyclic, hence the Hamiltonian,  
\ben
H = {\partial L \over \partial \dot{r} } \dot{r} - L 
  = \frac{R^{n-1}\triangle}{ 
                  \sqrt{ -\triangle + {\triangle}^{-1} {\dot{r}^2} } 
                            } \;,  
\label{eq:Hamiltonian}
\een
is conserved: $H=$constant$=: E$.   
Then, from (\ref{eq:Hamiltonian}), 
one obtains a first integral, 
\ben
 \left( \frac{1}{\triangle} {dr \over d t}\right)^2 
        - \frac{\triangle R^{2(n-1)}}{(E - \lambda F)^2} = 1 \;,   
\label{eq:first-integral}
\een 
which is analogous to the {\it law of energy conservation} for 
a point particle moving in an effective potential, 
\ben
 V_{\rm eff} := - \frac{\triangle R^{2(n-1)}}{(E - \lambda F)^2} \;.     
\een 

The geometry of $\Sigma$ is determined, once $(E,\lambda)$, and 
an ``initial'' point are given. 
The first and second fundamental forms of $\Sigma$ are written, 
in the coordinates $(r,z^p)$, as 
\bena
   h_{ij}dx^idx^j &=&\frac{1}{\triangle + \mu^2}dr^2 
                     + R^2 \Omega_{pq}dz^pdz^q \;, 
\\
   K^r_r &=& \lambda + (n-1) \mu \frac{R'}{R} \;, 
\label{Krr}
\\
   K^p_q &=& - \mu \frac{R'}{R} \delta^p_q \;,  
\label{Kpq}
\eena 
where $\mu(r) := \{E - \lambda F(r)\} R^{-(n-1)}$.

The qualitative behaviours of $\Sigma$ in $(M,g_{\mu \nu})$ 
can be understood by observing the shape of the function $V_{\rm eff}$. 
Inside the black hole, where $\triangle <0$, $V_{\rm eff}$ is positive and 
the extremal of $V_{\rm eff} $ provides a $r=$ constant hypersurface. 
Actually, for example in the Schwarzschild spacetime, 
one can obtain a $r=$ constant hypersurface for any $r(=R)< r_g$, 
by choosing $(E, \lambda)$ such that 
\ben
  E^2 = \frac{r^n(2r^{n-2}-nr_g^{n-2})^2}{4n^2(r_g^{n-2} - r^{n-2})} \;, 
\quad 
  \lambda^2 = 
    \frac{\{2(n-1)r^{n-1}-nr_g^{n-2}\}^2}{4r^n(r_g^{n-2} - r^{n-2})} \;, 
\label{parameters:E-Lambda}   
\een 
for which $V_{\rm eff} =1$ and $dV_{\rm eff}/dr =0$ are satisfied. 

On the other hand, outside the event horizon, $V_{\rm eff}$ is negative 
and there is no $r=$ constant hypersurface. If $\lambda \neq 0$, 
$V_{\rm eff} \rightarrow 0$ as $r \rightarrow \infty$,  
and then (\ref{eq:first-integral}) means $dr/dt \rightarrow 1$, that is, 
$\Sigma$ turns out to be asymptotically null and goes to null infinity. 
If $\lambda =0$, i.e., $\Sigma$ is maximal, then $\Sigma$ goes to 
spacelike infinity. 

Note that the solutions of (\ref{eq:first-integral}) do not involve 
$t=$ constant hypersurfaces. Such a hypersurface is time-symmetric, and 
thus totally geodesic, $K_{ij} =0$.

\begin{figure}[h] 
 \centerline{\epsfxsize = 9cm \epsfbox{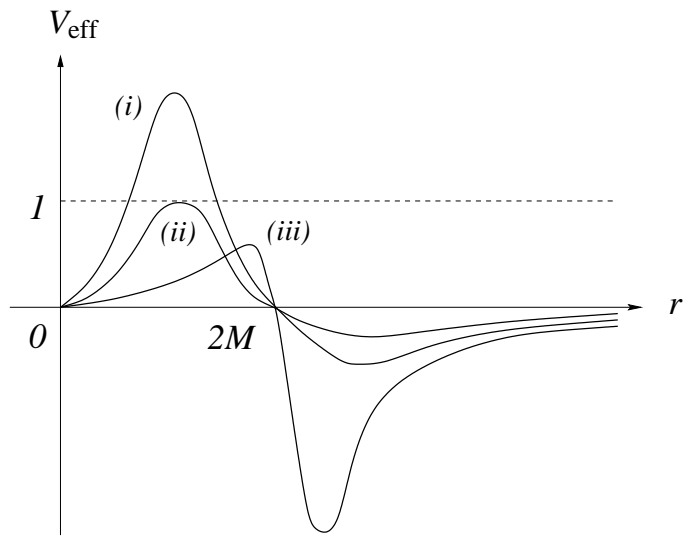}} 
\vspace{3mm}
        \caption{The effective potential $V_{\rm eff}$ of $\lambda
                 \neq 0$ case, which becomes zero at $r=0$ and $2M$ 
                 and $\sim - r^{-2}$ as $r \rightarrow \infty$. 
                {\sl (i)}: There are turning points. $\Sigma$ has 
                ether two singularities, or none at all.  
                {\sl (ii)}: There is an unstable equilibria point. 
                            $\Sigma$ is a $r=$ constant hypersurface.  
                {\sl (iii)}: There is no turning point. 
                            $\Sigma$ contains one and only one singularity.}  
        \protect 
        \label{fig:Potential} 
\end{figure}

\begin{figure}[h] 
 \centerline{\epsfxsize = 17.5cm \epsfbox{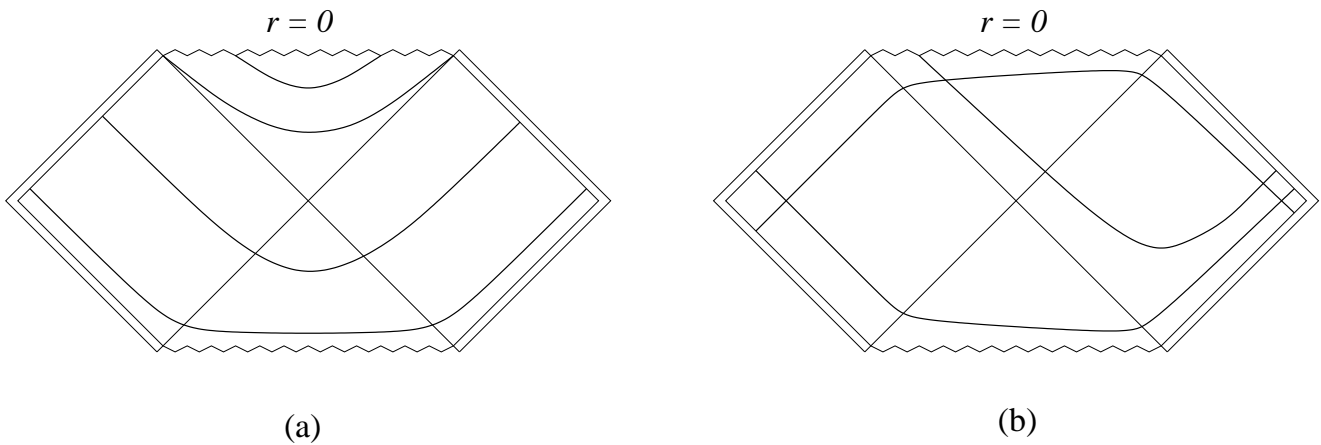}} 
\vspace{3mm}
        \caption{Penrose diagram of the Schwarzschild black hole and 
                 a variety of constant mean curvature hypersurfaces 
                 with various $(E,\lambda)$. 
                 (a) Hypersurfaces with reflection symmetry.
                 (b) No-reflection symmetric hypersurfaces.}
        \protect 
        \label{fig:CMCsurfaces} 
\end{figure}

\vs 
The numerical integration of (\ref{eq:first-integral}) 
has been done by Brill {\it et.al}~[BCI] 
in the $4$-dimensional ($n=3$) Schwarzschild black hole case, in which  
the effective potential becomes 
\ben
  V_{\rm eff} = - \frac{r^3(r-2M)}{\left(E-\frac{\lambda}{3}r^3
\right)^2} \;.  
\een 
The shape of the function $V_{\rm eff}$ and some typical 
hypersurfaces are schematically depicted in the
figures~\ref{fig:Potential}, and~\ref{fig:CMCsurfaces}, 
respectively~[BCI]. 

Some of the hypersurfaces which do not have a {\it turning point}, where 
$V_{\rm eff} =1$ or $dr/(\triangle dt) =0$, 
necessarily hit the singularity at $r=0$. 
It has been shown that for each fixed value of $\lambda$, 
there exit values $E_+$ and $E_-$ such that 
all hypersurfaces with $E< E_-$ or $E> E_+$ contain one and only one 
singularity, while those with $E_-<E< E_+ $ contain either two 
singularities or none at all. 
The value $E_\pm$ are determined by (\ref{parameters:E-Lambda}) and  
the hypersurfaces $E = E_+$ or $E = E_-$ are the homogeneous $r=$ constant 
hypersurfaces, which are the barriers for the regular hypersurfaces 
with a given $\lambda$. 
For example, in the maximal case $\lambda =0$, 
the barrier hypersurface is given by $r=3M/2$ 
and any regular maximal hypersurfaces do not exist 
inside this hypersurface. 

The point is that if $\lambda$ is large enough, regular constant mean 
curvature hypersurfaces can approach the singularity at $r=0$ 
arbitrarily closely. Thus, as discussed before, 
the regular ${\rm Tr} K = \lambda >0$ hypersurfaces can penetrate 
the barrier hypersurface $r=3M/2$. 
However a foliation by ${\rm Tr} K<0$ hypersurfaces, on each one of
which $r$ attains a minimum value cannot penetrate the barrier.

\vs

We should comment that ${\rm Tr} K>0$ does not always mean that $\Sigma$ is 
a convex hypersurface. 
Actually, in the present example, 
one of the components of the second fundamental form 
(\ref{Krr}) and (\ref{Kpq}) can be negative even when ${\rm Tr} K>0$. 
The existence of a spacetime convex function is thus a sufficient 
condition but not a necessary condition 
for the existence of a convex hypersurface. 

A particularly interesting case is that of $E=0$ with $r=R$, 
for which the second fundamental form becomes 
\ben
   K_{ij} = {\lambda \over n} h_{ij} \;, 
\een
hence $\Sigma$ is a totally umbilic convex hypersurface. 
In this case, as discussed in the previous section~\ref{sec:Levelsets}, 
one may find a convex function whose level coincides with $\Sigma$.

\vs 

We also remark that $\Sigma$ which has a turning point 
contains a minimal surface $S \subset \Sigma$. 
The second fundamental form $H_{pq}$ of $r=$ constant sphere 
$S \subset \Sigma$ is given by 
\ben 
   H_{pq} = - \sqrt{\mu^2(1 - V_{\rm eff})} \frac{R'}{R} h_{pq} \;. 
\een 
Thus $H_{pq} =0$ at the turning point. Hence $S$ is minimal there. 
This also means that $(\Sigma,h_{ij})$ contains closed geodesics on $S$.  
Since all regular hypersurfaces have a turning point, 
it turns out from Proposition~\ref{prop:no-closed submfd} that 

\begin{remark}
Regular constant mean curvature hypersurfaces in a black hole spacetime 
do not admit a strictly or uniformly convex function which lives on 
the hypersurface.  
\end {remark}

However, regular $\Sigma$s which foliate the interior of a black hole 
do not intersect the closed marginally inner and outer trapped surface.  
Thus, there remains a possibility that 
$(M,g_{\mu \nu})$ admits a {\sl spacetime convex} function whose level sets 
coincide with $\Sigma$s. 
%

\vs 

The existence problem of constant mean curvature foliations 
has been investigated extensively not only in black hole spacetimes 
as discussed here but also in cosmological spacetimes, or 
spacetimes having a compact Cauchy surface~(see, e.g. [Br,Gr,IR]). 
In cosmological spacetimes, constant mean curvature hypersurfaces, if exist, 
are likely to be compact, and thus do not admit a strictly or uniformly 
convex function which live on the hypersurfaces, because of 
Corollary~\ref{coro:cannot be closed}. 
However, a {\sl spacetime convex} function, if available, can give a 
constant mean curvature foliation with non-vanishing mean curvature 
as its level surfaces. 

\section{Conclusion}  
\label{sec:summary}

We have discussed consequences of the existence of convex functions on 
spacetimes. We first considered convex functions on a Riemannian 
manifold. We have shown that the existence of a suitably defined 
convex function is incompatible with closedness of the manifold and 
the existence of a closed minimal submanifold.  
%
%
We also discussed the relation to 
Killing symmetry and homothety.  
Next we gave the definition of a convex function on a 
Lorentzian manifold so that its Hessian has Lorentzian signature 
and the light cone defined by the Hessian lies inside the light cone 
defined by metric. Spacetimes admitting spacetime convex functions 
thus have a particular type of causal structure.  
We have shown that spacetime on which a spacetime convex function
exists does not admit a closed marginally inner and outer trapped
surface. We gave examples of spacetime convex functions on
cosmological spacetimes, anti-de-Sitter spacetime, and 
a black hole spacetime. Level sets of a convex function provide 
convex hypersurfaces. We have discussed level sets of 
convex functions, barriers, and foliations by constant mean curvature 
hypersurfaces. 
We anticipate that further study of convex functions and foliations by
convex surfaces will provide additional insights into global problems 
in general relativity and should have applications to numerical relativity.

\section*{\bf Acknowledgments}

G.W.G. would like to thank Profs. T. Nakamura and H. Kodama for their 
kind hospitality at the Yukawa Institute for Theoretical Physics 
where the main part of this work was done. 
A.I. would like to thank members of DAMTP for their kind hospitality. 
A.I. was supported by Japan Society for the Promotion of Science.


\end{document}